\documentclass[namedreferences]{solarphysics}
\usepackage[hyperref,optionalrh]{spr-sola-addons} 
\usepackage{graphicx}                    
\usepackage{color}                       
\usepackage{breakurl}                         

\usepackage{txfonts}
\usepackage{natbib}
\bibpunct{(}{)}{;}{a}{}{,}

\begin{document}
\begin{article}
\begin{opening}

\title{Impulsive Heating of Solar Flare Ribbons Above 10 MK}

\author{P.J.A.~\surname{Sim\~oes}$^{1}$ \sep D.R.~\surname{Graham}$^{1,2}$ \sep L.~\surname{Fletcher}$^{1}$}

\institute{
$^{1}$SUPA School of Physics and Astronomy, University of Glasgow, G12 8QQ, UK. 
email: \url{paulo.simoes@glasgow.ac.uk} email: \url{lyndsay.fletcher@glasgow.ac.uk} \\
$^{2}$current address: INAF-Osservatorio Astrofisico di Arcetri, I-50125 Firenze, Italy. 
email: \url{dgraham@arcetri.astro.it}
}

\runningtitle{Impulsive Heating of Flare Ribbons}
\runningauthor{Sim\~oes, Graham, Fletcher}

\date{Received / Accepted}

\begin{abstract}
The chromospheric response to the input of flare energy is marked by extended extreme ultraviolet (EUV) ribbons and hard X-ray (HXR) footpoints. These are usually explained as the result of heating and bremsstrahlung emission from accelerated electrons colliding in the dense chromospheric plasma. We present evidence of impulsive heating of flare ribbons above 10 MK in a two-ribbon flare. We analyse the impulsive phase of SOL2013-11-09T06:38, a C2.6 class event using data from \textit{Atmospheric Imaging Assembly} (AIA) on board of \textit{Solar Dynamics Observatory} (SDO) and the \textit{Reuven Ramaty High Energy Solar Spectroscopic Imager} (RHESSI) to derive the temperature, emission measure and differential emission measure of the flaring regions and investigate the evolution of the plasma in the flaring ribbons. The ribbons were visible at all SDO/AIA EUV/UV wavelengths, in particular, at 94 and 131 \AA\ filters, sensitive to temperatures of 8 MK and 12 MK. Time evolution of the emission measure of the plasma above 10 MK at the ribbons has a peak near the HXR peak time. The presence of hot plasma in the lower atmosphere is further confirmed by RHESSI imaging spectroscopy analysis, which shows resolved sources at 11--13 MK associated with at least one ribbon. We found that collisional beam heating can only marginally explain the necessary power to heat the 10 MK plasma at the ribbons.
\end{abstract}

\end{opening}

\section{Introduction}\label{sec:intro}

The impulsive phase of solar flares is characterised by intense emission of microwaves and hard X-rays (HXR), showing the presence of non-thermal electrons in the corona and chromosphere, respectively. These electrons are believed to be accelerated by the magnetic energy release mechanisms in the corona \citep[\textit{e.g.}][]{ZharkovaArznerBenz:2011} although this interpretation has been contested recently \citep{FletcherHudson:2008,BrownTurkmaniKontar:2009,VaradyKarlickyMoravec:2014}. Strong ultraviolet (UV) and extreme ultraviolet (EUV) emission is also commonly observed in association with the non-thermal emission \citep[\textit{e.g.}][]{HintereggerHall:1969,EmslieNoyes:1978,HoranKreplinFritz:1982,AlexanderCoyner:2006,CoynerAlexander:2009}, indicating fast heating at the transition region and chromosphere. This heating is usually explained by the energy deposition of the accelerated electrons hitting the high-density plasma at these locations \citep{Brown:1973,Hudson:1972,BrownKarlickyMacKinnon:1990}. The increase of pressure due to heating drives the plasma upwards into the coronal loops, a process termed chromospheric evaporation. The coronal loops filled with hot plasma will be bright in soft X-rays (SXR) and EUV, peaking at temperatures between 8 and 40 MK \citep[\textit{e.g.}][]{RyanMilliganGallagher:2012}. The maximum temperature is usually reached after the impulsive phase, \textit{i.e.} after most of the released energy is deposited in the ambient plasma. The hot coronal loops eventually cool by conduction, sending energy to the lower atmosphere, followed by a long radiative cooling phase \citep[\textit{e.g.}][]{Svestka:1987}.

 \cite{McTiernanKaneLoran:1993}, using \textit{Yohkoh} \textit{Soft X-ray Telescope} (SXT) images identified impulsive SXR emission associated with a HXR footpoint. Its duration was less than one minute, reaching its maximum about 20 seconds before an associated HXR sub-burst was detected by the low energy channel, and 40 seconds before the main HXR burst at the same location. This SXR burst corresponded to a sharp temperature-time profile, rising from $\approx 7$ MK to about $\approx 10$ MK in $\approx 30$ seconds, followed by a fast cooling. \cite{HudsonStrongDennis:1994} reported similar observations for the limb event SOL1992-01-26, where the footpoint SXR peak time matched the HXR peak time within the instrument's temporal resolution. \cite{MrozekTomczak:2004} studied SXR impulsive brightenings in footpoints in 46 \textit{Yohkok} events, and concluded that these can be explained as result of collisional heating mainly by non-thermal electrons with relatively low energies. 

More recently, \cite{GrahamHannahFletcher:2013} presented the emission measure distributions (EMD) as a function of temperature for the footpoints of six flares, derived from EUV spectroscopic observations, and showed that the footpoint EMDs had a peak around 8 MK, indicating the presence of high temperature plasma within the footpoints. Evidence of hot ($T>1$ MK) and dense ($n_e>10^{10}~\mathrm{cm}^{-3}$) plasma at flare footpoints during the impulsive phase was also presented by several other authors \citep{MilliganDennis:2009,WatanabeHaraSterling:2010,Del-ZannaMitra-KraevBradshaw:2011,Milligan:2011,GrahamFletcherHannah:2011,FletcherHannahHudson:2013}. A comprehensive review of EUV spectroscopic observations of hot footpoints has been presented by \cite{Milligan:2015}.

While the \textit{Hinode}/{\em Extreme ultraviolet Imaging Spectrometer} \citep[EIS:][]{CulhaneHarraJames:2007} provides great spectroscopic information on the plasma across a wide temperature range, the slit-raster observational technique does not allow evaluation of the plasma evolution at the same spatial location with high temporal cadence. Imaging from the {\em Atmospheric Imager Assembly} \citep[AIA:][]{LemenTitleAkin:2012}, on board the {\em Solar Dynamics Observatory} (SDO), and methods to recover the EMDs allow us to track the plasma evolution at different regions of flares simultaneously. We present the analysis of the SOL2013-11-09 flare ribbons, recovering the temperature and emission measure and show evidence for a sudden heating to $\approx 10$ MK temperatures of the ribbon plasma during the impulsive phase. 

\section{Observational Data of SOL2013-11-09}\label{sec:overview}

The SOL2013-11-09 flare occurred in active region NOAA 11890 near disc centre, and was characterised by two flaring ribbons plus a bright compact source between them during the impulsive phase. For this study we used data from SDO/AIA. AIA has nine filters, or passbands, in the EUV and UV range: 94, 131, 171, 193, 211, 304, 335, 1600 and 1700 \AA~ filters, taking full Sun images with a pixel resolution of 0.6 arcsecs and 12 seconds cadence for the EUV passbands and 24 seconds for the UV passbands. We also used data from the {\em Reuven Ramaty High-Energy Solar Spectroscopic Imager} \citep[RHESSI:][]{LinDennisHurford:2002}. The RHESSI imager consists of nine bi-grid rotating modulation collimators (RMCs), with a rotation period of about 4 seconds. The data from individual RMC are combined by the imaging algorithms to cover different spatial scales, from 2.3 arcsecs (RMC 1) up to 180 arcsecs (RMC 9). To avoid saturation of the detectors at times with high rate of incident photons, two-stage (``thin" and ``thick") attenuators can be automatically employed. No attenuators (state ``A0") were in during this event, a rare occurrence that gives us the opportunity to investigate the evolution of the low-energy X-rays during the impulsive phase without gaps and with full count rates. Complementary SXR data from the {\em Soft X-ray Sensor} on board of the {\em Geostationary Operational Environmental Satellite} (GOES) were also employed. GOES observes the Sun as a star in two broadband SXR channels, 1--8 \AA~ and 0.5--4 \AA, which can be used to estimate the temperature, $T$, and emission measure, EM, of the flaring plasma \citep{ThomasCrannellStarr:1985,WhiteThomasSchwartz:2005}.

\begin{figure} 
 \centerline{\includegraphics[angle=0,width=\textwidth]{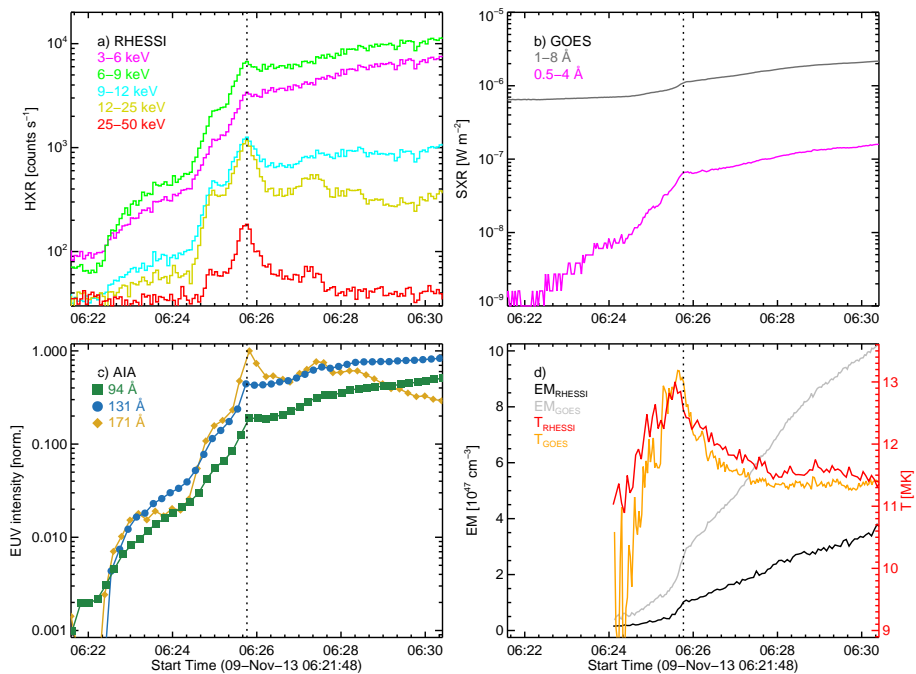}}
 \caption{SOL2013-11-09 flare overview. (a) RHESSI HXR count rates (energy bands indicated in the figure). (b) GOES flux at 1--8 and 0.5--4 \AA \ channels. (c) Spatially integrated emission of AIA passbands 94, 131, 171 \AA. {\em Bottom:} (d) Temperature, $T$, and emission measure, EM, determined using from GOES and RHESSI. In all frames, the vertical dotted line indicates the HXR peak.}
 \label{fig:overview}
 \end{figure}

In Figure \ref{fig:overview} we summarise some aspects of the event. Figure \ref{fig:overview}a shows RHESSI HXR count rates at several energy bands, where the high-energy bands (9--12, 12--25 and 25--50 keV) show a clear impulsive peak with a maximum at 06:25:46~UT, and are associated with bremsstrahlung emission from non-thermal electrons, as we discuss in Section \ref{sec:rhessi}. A second, weaker and less impulsive peak is seen around 06:27:20~UT. The lower energy bands (below 9 keV), associated with thermal bremsstrahlung, show a gradual rise starting slightly after 06:22~UT, and also a pronounced peak simultaneous with the high-energy channels. Both GOES channels, 1--8 \AA~ and 0.5--4 \AA~ (Figure \ref{fig:overview}b) have a steep rise, with a ``shoulder'' at the same time as the HXR peak. Spatially integrated EUV emission was obtained by summing the pixels of AIA images taking the whole flare region and are shown in Figure \ref{fig:overview}c. The 171 \AA~emission is well associated with the high-energy HXR, with a main impulsive peak around 06:25:46~UT and a secondary peak around 06:27:20~UT. Similar characteristics are observed at 193, 211, 304, 1600, and 1700 \AA. Emission in the 94 and 131 \AA\ passbands is similar to the GOES SXR channels: a sharp rise matching the HXR peak, followed by a more gradual rise. These AIA passbands are sensitive to emission from hot plasma: the main contribution to the 94 \AA~ passband is the Fe {\sc xviii} 93.93 \AA\ emission line, formed at $\log T=6.8$ while the dominant contribution to the AIA 131 \AA\ passband during flares comes from Fe {\sc xxi} $128.75$ \AA~ ($\log T = 7.1$) \citep{ODwyerDel-ZannaMason:2010,BoernerTestaWarren:2014}. These main contributions were verified using from the \textit{EUV Variability Experiment} \citep[EVE:][]{WoodsEparvierHock:2012} data for this event by \cite{Simoes:2015}.
 
\section{SDO/AIA Data Analysis} \label{sec:aia}

\subsection{Impulsive EUV Emission from Flaring Ribbons} \label{sec:dem}

\begin{figure} 
\centerline{\includegraphics[angle=0,width=\textwidth]{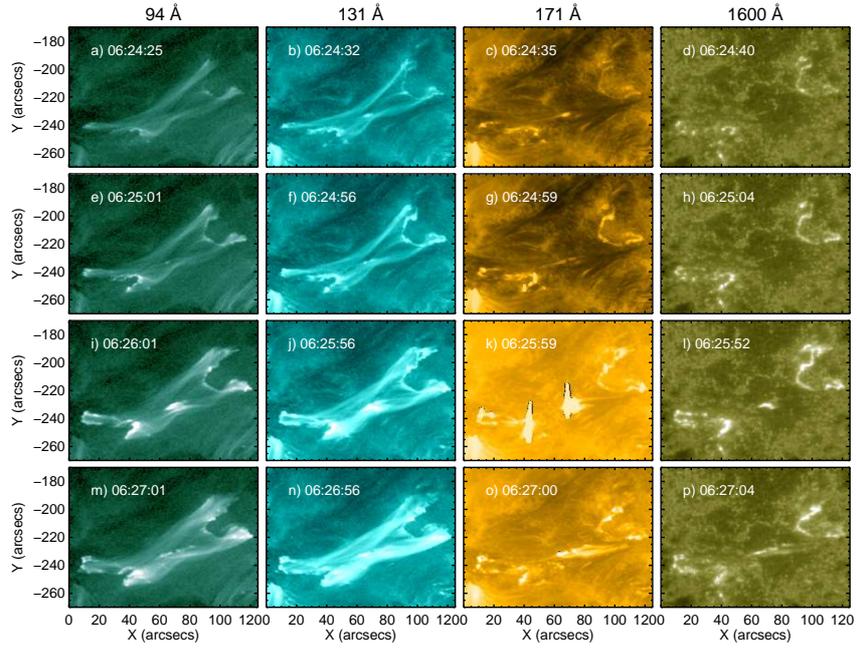}}
\caption{SDO/AIA images at 94, 131, 171 and 1600 \AA\ filters of SOL2013-11-09, at four times during the impulsive phase.}
\label{fig:aia}
\end{figure}

\begin{figure} 
\centerline{\includegraphics[angle=0,width=\textwidth]{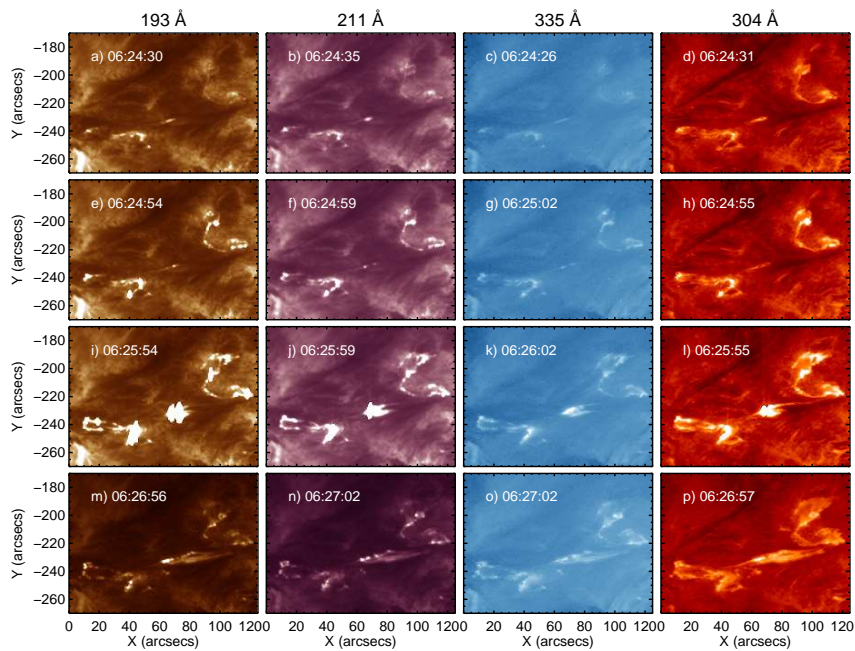}}
\caption{SDO/AIA images at 193, 211, 335 and 304 \AA\ filters of SOL2013-11-09, at four times during the impulsive phase.}
\label{fig:aia2}
\end{figure}

\begin{figure} 
\centerline{\includegraphics[angle=0,width=\textwidth]{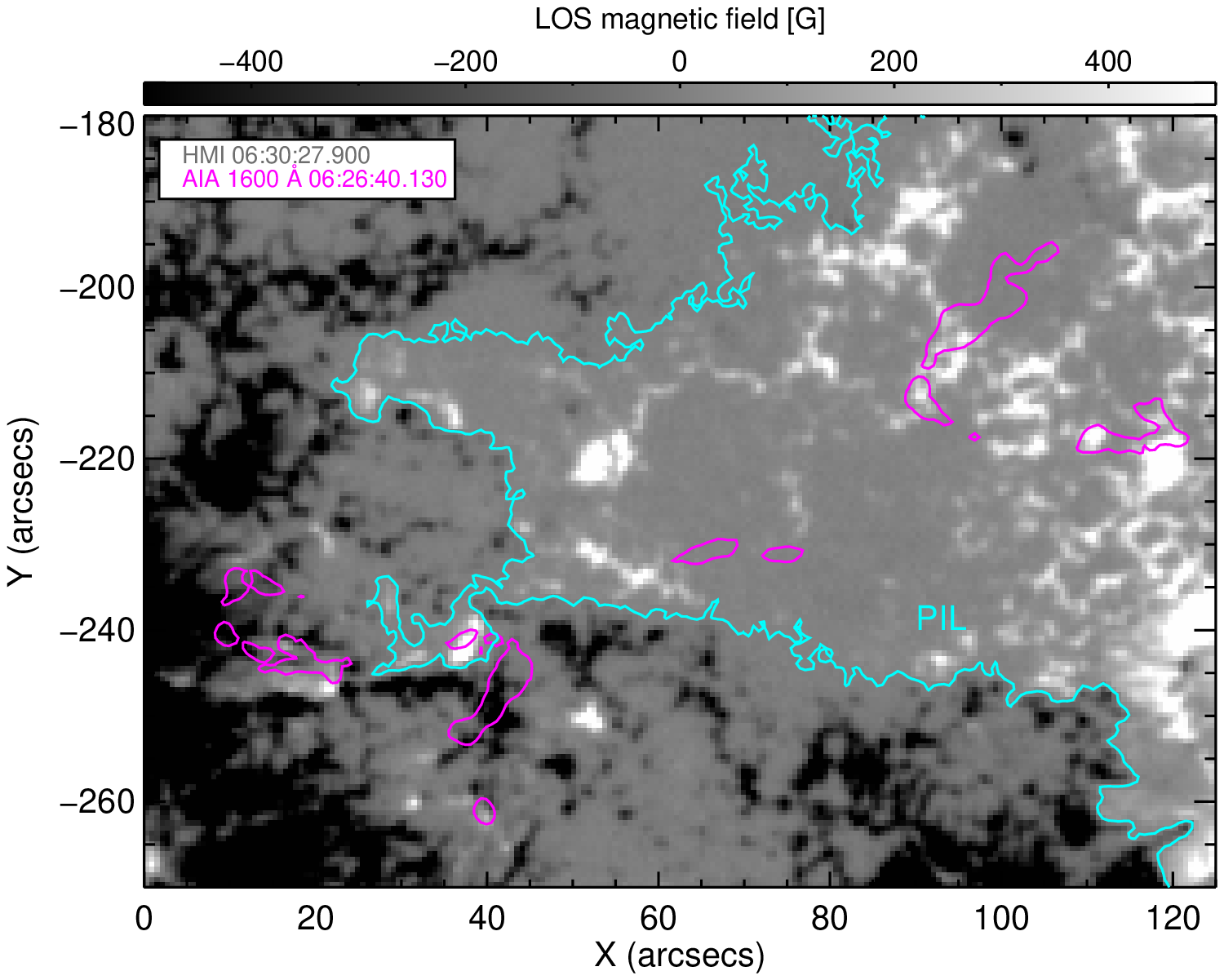}}
\caption{SDO/HMI line-of-sight (LOS) magnetogram, overlaid by SDO/AIA 1600 \AA\ contours at 500 DN s$^{-1}$, showing the position of the ribbons and the coronal source, analysed by \cite{Simoes:2015}.}
\label{fig:hmi}
\end{figure}

We will now focus our analysis on the impulsive phase, 06:22 to 06:30~UT, when the impulsive ribbon emission is observed. AIA images { (see Figure \ref{fig:aia} and \ref{fig:aia2})} reveal two ribbons as the brightest features in all AIA EUV/UV passbands during the impulsive phase of the event. { The ribbons are located in regions with opposite magnetic field polarity, as can be seen in Figure \ref{fig:hmi}, which shows the line-of-sight (LOS) photospheric magnetic field obtained by SDO/\textit{Helioseismic and Magnetic Imager} \citep[HMI:][]{Scherrer:2012}.} Coronal loops become the dominant features at 94, 131 and 335 \AA \ after the main impulsive peak of the flare. The bright source located between the ribbons, visible at the impulsive phase in all AIA wavelengths and also seen in HXR (Section \ref{sec:rhessi}), was analysed in detail by \cite{Simoes:2015}. They characterised this source as the main energy release site of this event. We summarise their findings as follows: 1) the source is located in the corona (possibly low down), it has a filamentary shape along the loops seen in EUV/UV images, the lack of hot loops connecting the region after the impulsive phase, and the weak and featureless photospheric magnetic field at the same location (see Figure \ref{fig:hmi}). 2) it has intense and impulsive EUV and HXR emission. 3) consistently, the source is found to be dense ($\log n=11.50\pm 0.82$) and hot ($T \approx 12\sim16$ MK). 4) strong red-shifts observed in many EUV emission lines observed by \textit{Hinode}/EIS, including Fe {\sc xii} and Fe {\sc xxiv}, indicate plasma downflows of 40--250 km s$^{-1}$, which, along with plasma outflows observed in AIA images, are interpreted as plasma outflows along the magnetic loops. 

\begin{figure} 
\centerline{\includegraphics[width=\textwidth,angle=0]{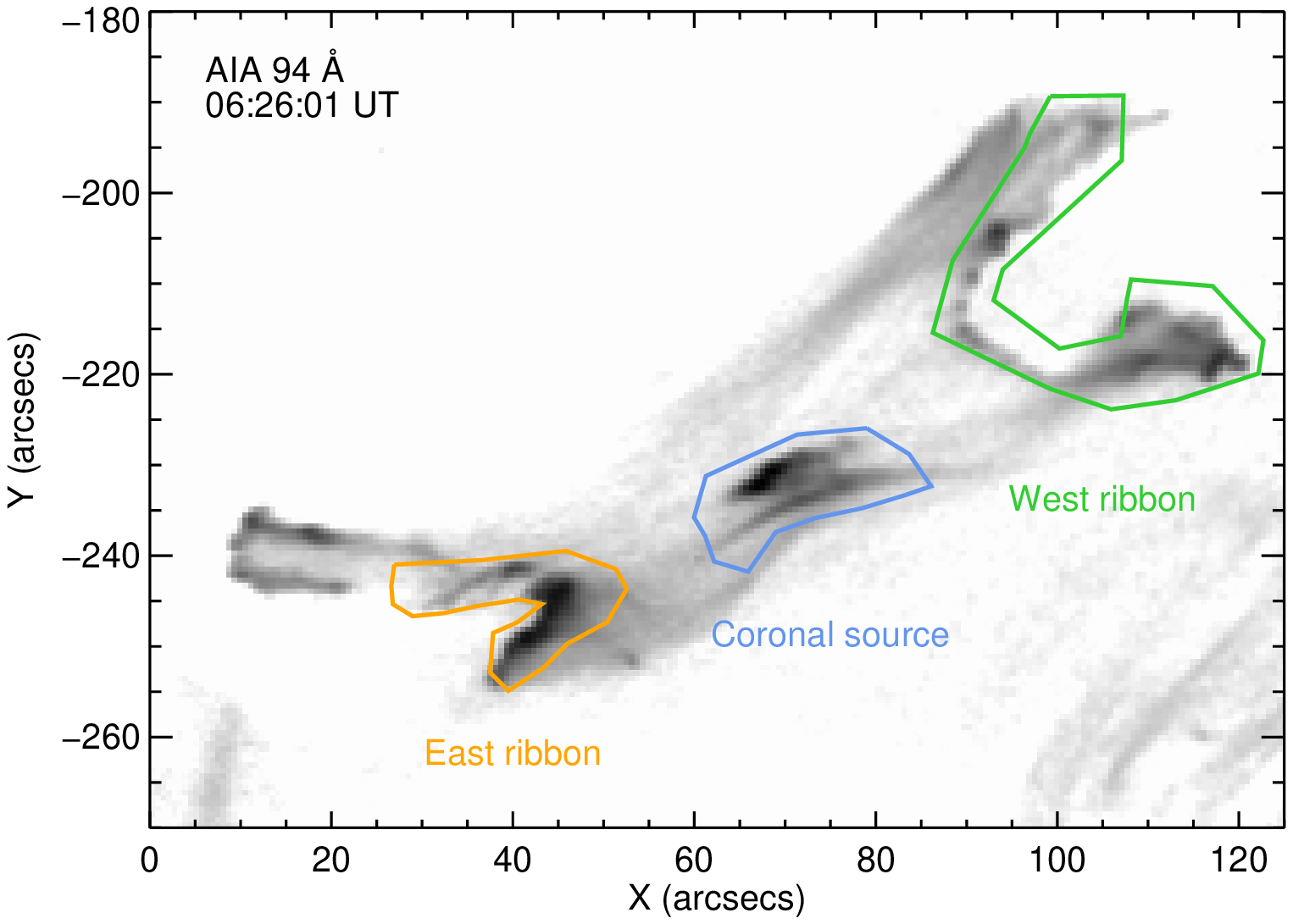}}
\caption{a) Regions of interest (ROI) defined for the AIA images: East (orange) and West (green) ribbons, and coronal source (blue).}
\label{fig:aia_roi}
\end{figure}

\begin{figure}  
\centerline{\includegraphics[angle=0,width=\textwidth]{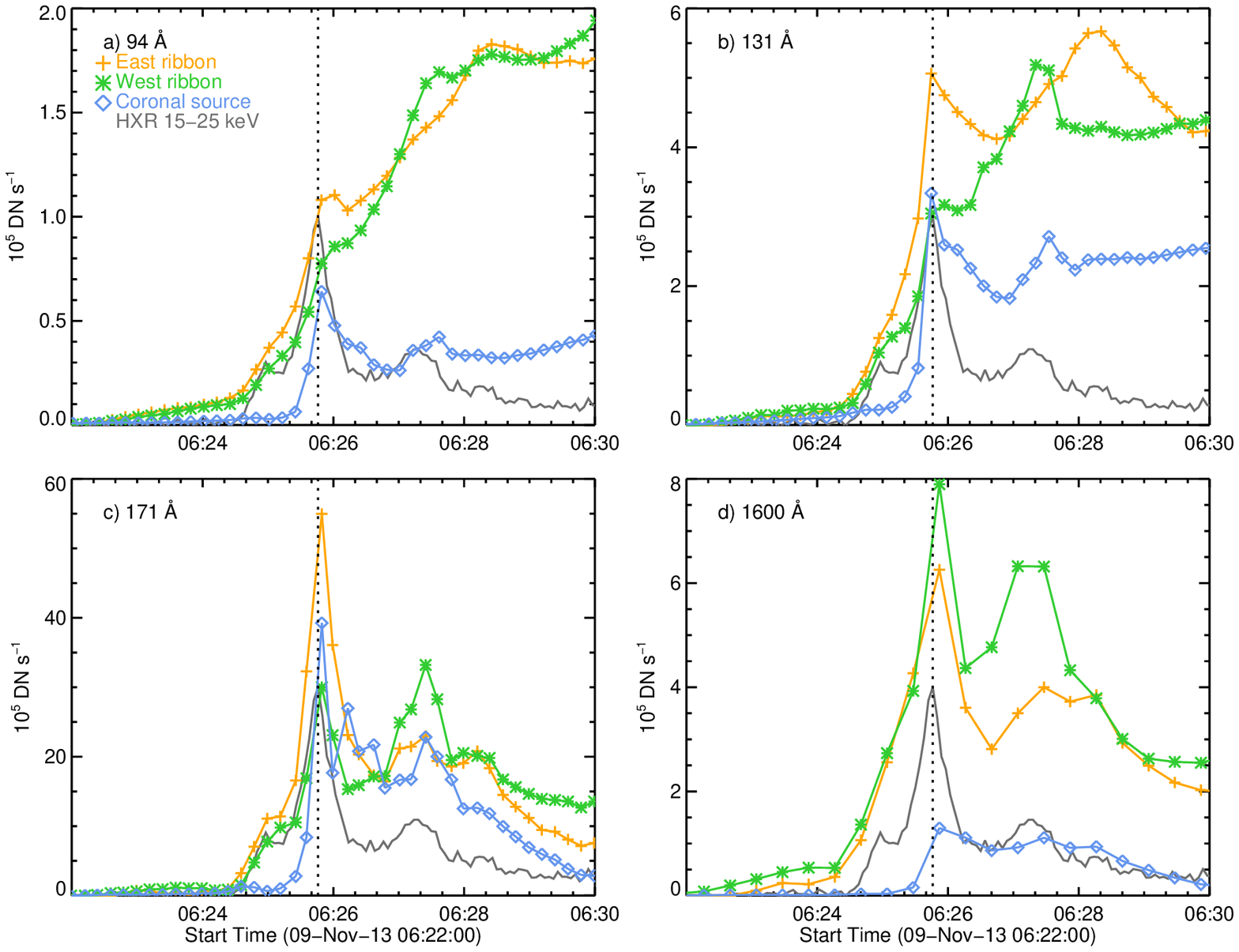}}
\caption{SDO/AIA lightcurves for EUV/UV filters, (a) 94 \AA, (b) 131 \AA, (c) 171 \AA\ and (d) 1600 \AA, for each region of interest (ROI), as defined in Figure \ref{fig:aia_roi}: East (orange) and West (green) ribbons, and coronal source (blue). RHESSI 15-25 keV counts (grey) are shown as a reference of the impulsive non-thermal emission, with the time of its maximum shown by the vertical dotted line.}
\label{fig:aia_peaks} 
\end{figure}

We investigate the emission from the ribbons and the coronal source (for context) separately by marking regions of interest (ROI) associated with each of the three main sources as shown in Figure \ref{fig:aia_roi}: East (orange) and West (green) ribbons, coronal source (blue). We obtained lightcurves for all AIA passbands by summing up the pixels for each ROI, which are shown in Figure \ref{fig:aia_peaks}. We also show RHESSI HXR 15-25 keV counts for comparison. The UV (1600 and 1700 \AA), 304 \AA, and the ``warm'' EUV (171, 193, 211 \AA) channels have similar enhancements during the impulsive phase, showing impulsive emission from both ribbons and the coronal source. The bursts are coincident with the HXR main peak (06:25:46~UT), within the instrumental cadence. A secondary, less impulsive HXR peak at 06:27:14~UT is also well-associated with most AIA channels. The peaks are simultaneous at the three ROIs in all nine AIA filters, within the instrumental cadence. After about 06:30~UT the coronal source fades out at all wavelengths. At 94 and 131 \AA \ the coronal source ROI is dominated by the emission from the coronal loops. 

\subsection{Emission Measure Distributions}

To investigate the evolution of the plasma at the flaring ribbons, we apply a method of regularised inversion to the AIA data developed by \cite{HannahKontar:2012} to obtain the {\em differential emission measure} (DEM), $\xi(T)={n_e}^2\mathrm{d}h/\mathrm{d}T$ $[\mathrm{cm}^{-5}$ K$^{-1}]$, defining the quantity of emitting material as a function of temperature, $T$, along the given line-of-sight, $h$, with an average density $n_e$. We employed up-to-date AIA temperature response functions \citep{BoernerEdwardsLemen:2012,BoernerTestaWarren:2014} which have empirical corrections for the missing emission lines from the CHIANTI v7.1.3 database \citep{DereLandiMason:1997,LandiYoungDere:2013}, time-dependent response corrections for each channel due to degradation of the filters, and normalisation to ensure agreement with SDO/EVE spectroscopic full-disk observations. The regularisation method also provides the DEM uncertainty and effective temperature resolution. To obtain the DEM it must be assumed that the emitting plasma is optically thin, in local thermodynamic equilibrium, and in ionisation equilibrium. In the case of a flare ribbon, where the emitting plasma is both hot and dense, this may be hold, although departures from equilibrium and optical depth effects should never be discounted completely \cite[see][for further discussion]{GrahamHannahFletcher:2013}. Continuum emission may also contribute to the AIA passbands \citep{ODwyerDel-ZannaMason:2010}, however this contribution is likely to be small for C class flares and can probably be neglected \citep{MilliganMcElroy:2013}. We point the reader to \cite{HannahKontar:2012} for an extended explanation of the method and comparison with other available methods.

\begin{figure}  
\centerline{\includegraphics[angle=0,width=\textwidth]{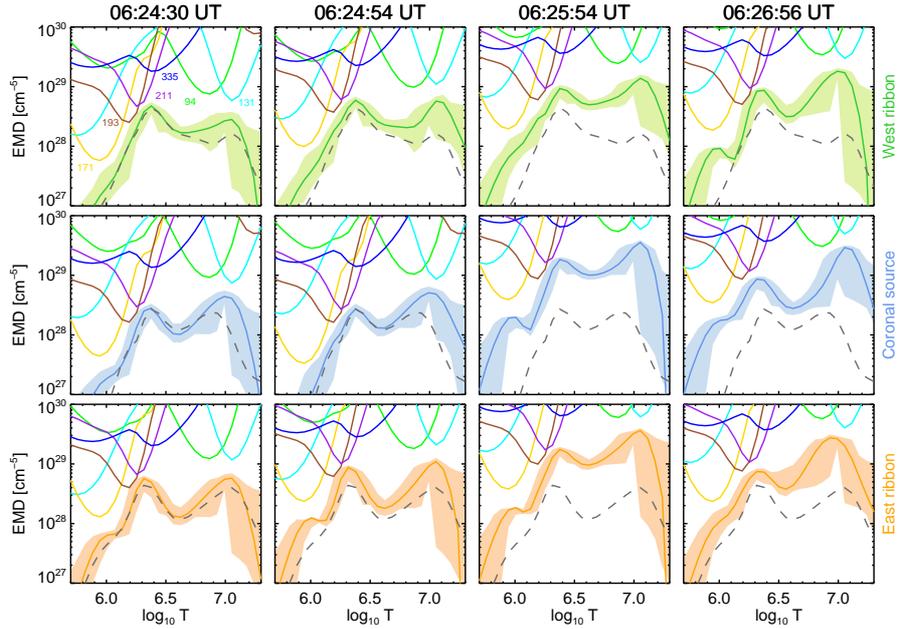}}
\caption{Emission measure distribution $\mathrm{EMD}$ (cm$^{-5}$) for the East (orange) and West (green) ribbons and coronal source (blue) at selected times from the early phase into the impulsive phase. Pre-flare (06:20:19~UT) EMD are shown by the grey dashed line for comparison. { The EM-\textit{loci} curves for each filter are indicated by different colours and identified in the top-left panel.} }
\label{fig:emds}
\end{figure}

We applied the method to the average emission inside each ROI indicated in Figure \ref{fig:aia_roi} throughout the impulsive phase. Integrating the DEM over fixed temperature intervals gives the emission measure distribution, $\mathrm{EMD}=n^2h$, in the more practical units of cm$^{-5}$ \cite[\textit{e.g.}][]{GrahamHannahFletcher:2013,HannahKontar:2013}. The resulting $\mathrm{EMD}$ for the East and West ribbons and coronal source for selected times are shown in Figure \ref{fig:emds} along with the EM-{\em loci} curves for each filter, these representing the $\mathrm{EMD}$ for an isothermal plasma, i.e the maximum theoretical EM.

The pre-flare EMD is shown by the dashed line with a peak around $\log T = 6.3 \sim 6.4$ and falling off towards higher temperatures. A second peak is visible around $\log T=6.9 \sim 7.1$, most notable in the East ribbon, however the 94 and 131~\AA\ AIA channels both contain contributions from lines at approximately 1 MK and 10 MK, therefore it is likely that this apparent hot emission is in fact much cooler. The $\mathrm{EMD}$ for the three sources share a similar development; a pre-flare EMD which evolves with an overall increase of the emission measure at all temperatures in the AIA passband response range (roughly $5.7 < \log T < 7.3$), a stronger increase in high temperature emission at $\log T \simeq 7.0$, and the formation of a low temperature ``shoulder'' just below $\log T = 6.0$.

\begin{figure}  
\centerline{\includegraphics[angle=0,width=\textwidth]{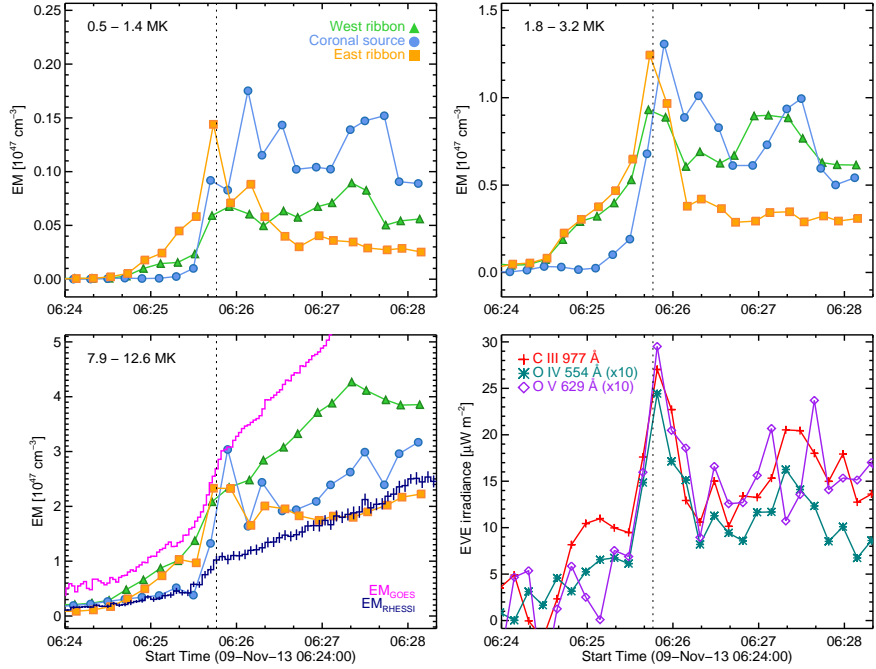}}
\caption{Time evolution of the EM (converted to $\mathrm{cm}^{-3}$ for comparison with GOES and RHESSI results) at three temperature ranges, (a) 0.5--1.4, (b) 1.8--3.2, and (c) 7.9--12.6 MK, for East ribbon, West ribbon, coronal source from AIA from pre-flare into the impulsive phase. (d) SDO/EVE lightcurves of selected emission lines formed at transition region temperature ($\log T \approx 5$). The vertical dotted line indicates the time of the HXR peak.}
\label{fig:em_time}
\end{figure}

We then integrated the $\mathrm{EMD}$ at temperature ranges 0.5--1.5 MK, 1.8--3.2 MK and 7.9--12.6 MK, which show the most prominent peaks, to obtain the column emission measure, $\mathrm{EM}_c=n^2h$~$[\mathrm{cm}^{-5}]$, at these temperature ranges. We did not attempt this at the temperatures around $\log T= 6.6$, as this range is not very well constrained by the AIA filters. In order to compare these results with (volume), $\mathrm{EM}=n^2V~[\mathrm{cm}^{-3}]$, values from GOES and RHESSI, we converted the $\mathrm{EM}_c$ to $\mathrm{EM}$ by multiplying the values by the projected area, $A$, of the sources, \textit{i.e.} making $V=hA$. The projected area $A$ of the emitting plasma was estimated by finding the number of pixels inside each ROI with a value above a determined threshold. Here we used 94 \AA\ images and chose a threshold value of 35 DN s$^{-1}$ pixel$^{-1}$ that captures the weaker but relevant emission at the early phase, and also the stronger emission at the impulsive phase. We verified if the emitting pixels at the other AIA filters sensitive to cooler plasma temperatures co-aligned well with the emitting area at 94 \AA, generally finding a good agreement within a couple of pixels. The time evolution of the EM (now in cm$^{-3}$) at the three temperature ranges for the three sources is shown in Figure \ref{fig:em_time}, along with the values from GOES and RHESSI obtained for full-Sun observations. We subtracted the EM pre-flare values to consider the flare excess only, taking the pre-flare values as the mean in the interval 6:00--6:20~UT. Looking at Figures \ref{fig:em_time}a-c in the early phase (6:22:00--6:25:20~UT), the EM for the two ribbons in all three temperature bands starts to rise very slowly from about 6:22~UT (clearly visible on a logarithmic scale, not shown here), with a faster enhancement after 6:24:40~UT. During the main impulsive phase (6:25:20--6:26:40~UT) the EM in the three temperature ranges for all the three sources rises impulsively, peaking around 6:25:40--6:26:20~UT. The GOES and RHESSI emission measures also show a steeper enhancement. This phase coincides with the main HXR peak. The rapid increase, followed by a decrease of the EM implies fast heating and cooling (or removal) of the plasma, and it is more pronounced at the East ribbon. After 6:26:40~UT, at 7.9--12.6 MK, the EM keeps rising, which is consistent with loop filling by chromospheric evaporation. AIA images at 94, 131 and 335 \AA \ show the enhancement of bright loops in this phase, and the increase is also supported by the GOES and RHESSI EM values.

At lower temperature ranges, the EM peak is more pronounced, showing a fast increase and decrease in the amount of material below $\approx 3$ MK. This behavior is similar to emission lines observed by SDO/EVE formed at chromospheric and transition region temperatures, $4.2 < \log T < 5.7$K, which show impulsive emission in time with HXR. In Figure \ref{fig:em_time}d we show the lightcurves of C {\sc iii} 977 \AA, O {\sc iv} 554 \AA~and O {\sc v} 629 \AA, noting that other lines formed at similar temperatures display similar behaviour, namely Ne {\sc vii} 465 \AA, O {\sc iii} 526 \AA, He {\sc i} 584 \AA, O {\sc iii} 599 \AA, O {\sc iv} 790 \AA, O {\sc vi} 1032 \AA, with some being noisier and less impulsive. Simulations by \cite{FisherCanfieldMcClymont:1985} show that the entire flare chromosphere cools rapidly on a radiative timescale after the heating is turned off. However, in this flare, after the main HXR peak there is still HXR and EUV emission present, indicating that the energy deposition is still occurring.

\section{RHESSI Hard X-rays Observations}\label{sec:rhessi}

\begin{figure} 
\centerline{\includegraphics[angle=0,width=0.95\textwidth]{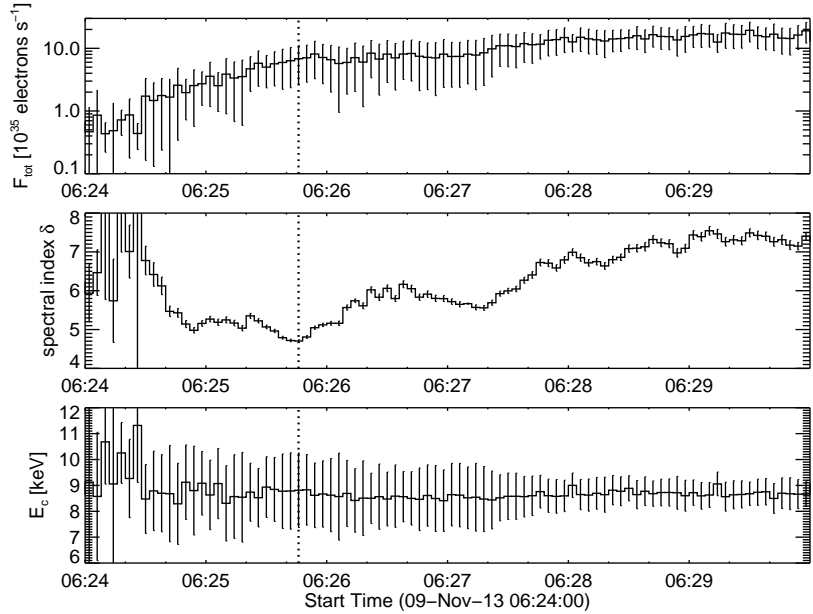}}
\caption{Spatially unresolved RHESSI spectral analysis results for the non-thermal electrons: $F_\mathrm{tot}$, the total electron rate above the low energy cutoff $E_c$ and spectral index $\delta$.}
\label{fig:rhessi}
\end{figure}

We now employ RHESSI HXR observations to characterise the hot plasma and the non-thermal electrons. First, we evaluate the HXR spectra during the impulsive phase, with a 4 second time resolution. RHESSI spectra were fitted with an isothermal plus a single power-law thick-target model, using \textit{Object Spectral Executive} (OSPEX) software \citep{SchwartzCsillaghyTolbert:2002}. No attenuator was active during the course of the event (state A0). The non-thermal power-law is assumed to originate from a cold, collisional thick-target \citep{Brown:1971}. The fitting parameters for the non-thermal electrons derived from RHESSI are shown in Figure \ref{fig:rhessi}. We will use these parameters to derive the collisional beam heating in Section \ref{sec:heating}. The thermal plasma parameters (EM and $T$) are shown in Figure \ref{fig:overview} along with the same parameters estimated from GOES data.

\begin{figure} 
\centerline{\includegraphics[angle=0,width=0.99\textwidth]{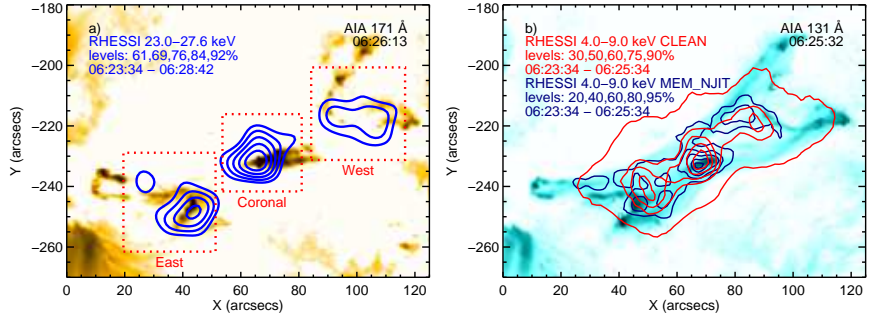}}
\caption{(a) RHESSI HXR 23--27 keV contours overlaid on an SDO/AIA 171 \AA\ image (inverted colours) with (red) dashed boxes indicating the regions for imaging spectroscopy. (b) RHESSI HXR 4--9 keV contours obtained with Clean (red) and MEM NJIT (dark blue) overlaid on an SDO/AIA 131 \AA~ image (inverted colours).}
\label{fig:imsp_roi}
\end{figure}

\begin{figure} 
\centerline{\includegraphics[angle=0,width=0.99\textwidth]{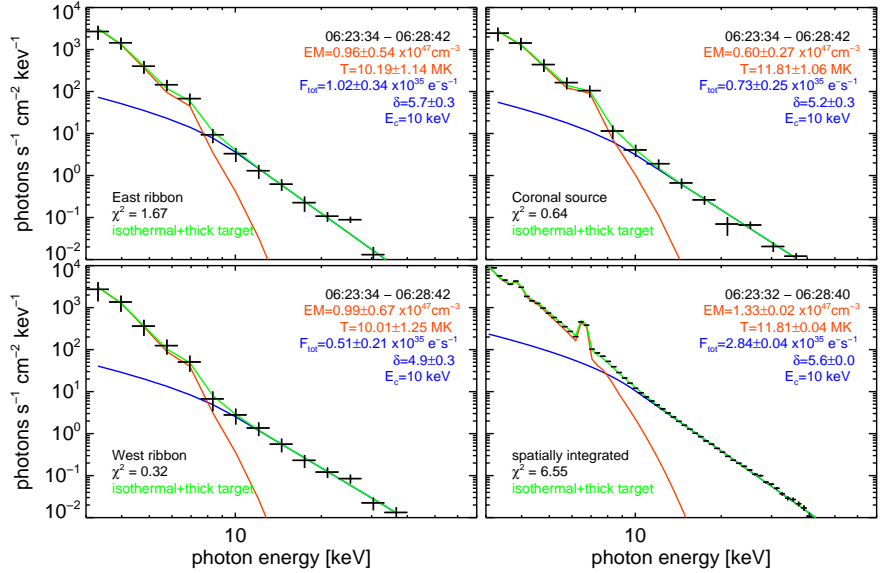}}
\caption{RHESSI photon spectra for the three main HXR sources indicated in Figure \ref{fig:imsp_roi} and spatially integrated, for the interval 06:23:34 -- 06:26:42~UT.}
\label{fig:imsp}
\end{figure}

We applied the imaging spectroscopy technique for the three main HXR sources observed associated with the EUV ribbons and coronal source, indicated in Figure \ref{fig:imsp_roi}. Due to RHESSI dynamic range, it was not possible to do imaging spectroscopy for the West ribbon. We constructed RHESSI Clean images for 14 energy bands, logarithmically spaced between 3 and 40 keV, using detectors 3 to 8, integrated during the main impulsive phase 06:23:34 -- 06:26:42~UT. The spectra were fitted with an isothermal plus thick-target model, shown in Figure \ref{fig:imsp}, and the resulting parameters are found in Table \ref{tab:imsp}. For comparison, we also fitted the spatially unresolved full Sun spectrum integrated for the same time interval. The thermal and non-thermal properties of the three HXR sources are very similar, confirming that the ribbon plasma is above 10 MK during the impulsive phase, as indicated by our AIA analysis in Section \ref{sec:dem}. Also, the non-thermal properties are in agreement with the findings of \cite{Simoes:2015} { who showed that the coronal source is 9--18 arcsecs long, with a plasma density of $\log n=11.5$ and thus being collisionally thick for electrons with energies of up to 45--65 keV.}

\begin{table}
\caption{RHESSI imaging spectroscopy results.}
\label{tab:imsp}
\begin{tabular}{lllll}
\hline
& EM $\times 10^{47}\mathrm{cm}^{-3}$ & $T$[MK] & $F_{35} 10^{35} s^{-1}$ & $\delta$ \\
\hline
East ribbon &$1.0 \pm 0.5$ & $10 \pm 1 $&$ 1.0\pm 0.3 $&$ 5.7 \pm 0.3$ \\
Coronal &$0.6 \pm 0.3$ & $12 \pm 1 $&$ 0.7 \pm 0.2 $&$ 5.2 \pm 0.3$ \\
West ribbon&$1.0 \pm 0.7$ & $10 \pm 1 $&$ 0.5 \pm 0.2 $&$ 4.9 \pm 0.3$ \\
Full Sun &$1.33 \pm 0.02$ & $11.81 \pm 0.05 $&$ 2.84 \pm 0.04$ &$ 5.61 \pm 0.02$ \\
\hline
\end{tabular}
\end{table}

\section{Discussion}\label{sec:discussion}
\subsection{Impulsive Heating of Ribbons}

\begin{figure} 
\centerline{\includegraphics[angle=0,width=\textwidth]{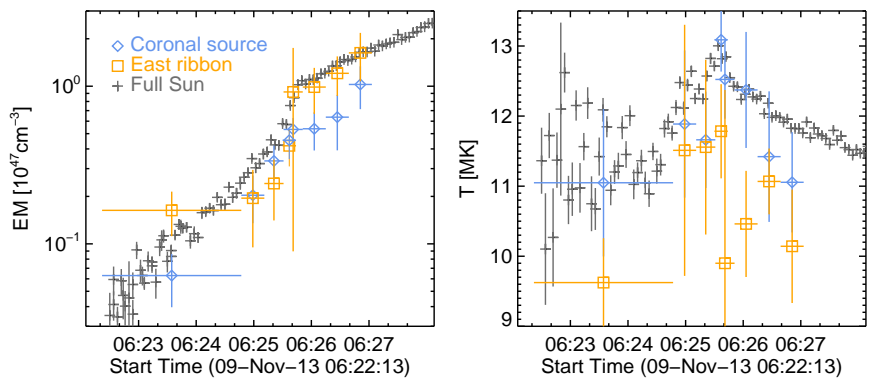}}
\caption{EM and T from RHESSI imaging spectroscopy for the coronal source
  (blue crosses) and East ribbon source (orange). The EM and $T$ from the unresolved
  spectral fitting are also shown (grey).}
\label{fig:spot_em_t}
\end{figure}

\begin{figure} 
\centerline{\includegraphics[angle=0,width=\textwidth]{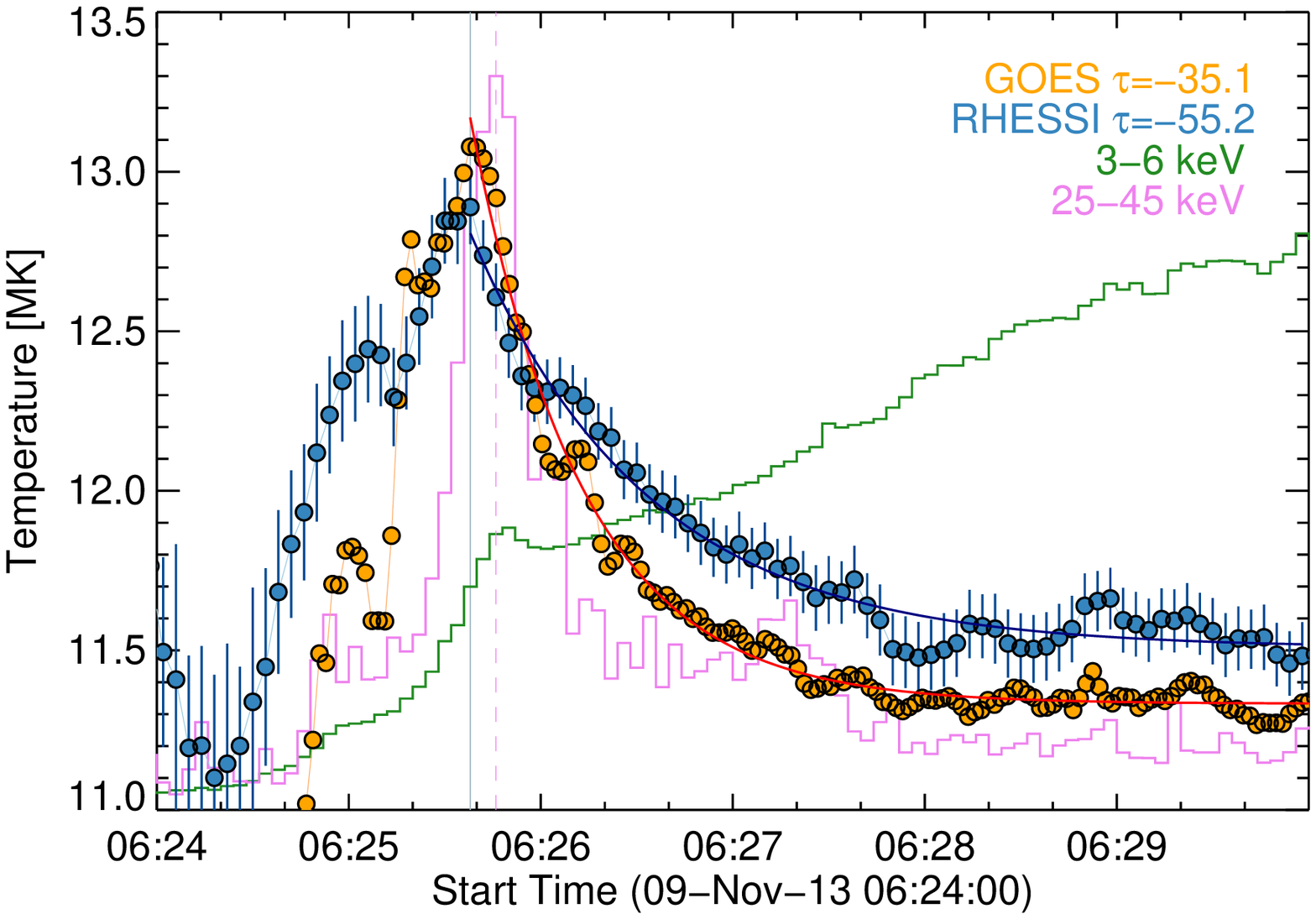}}
\caption{Plasma temperature inferred from RHESSI (blue) and GOES (orange) data, peaking about 10 seconds before the RHESSI HXR counts at 3--6 keV (green) and 25--45 keV (violet). The values of the $e$-folding cooling time, $\tau$ (in seconds), are indicated in the figure.}
\label{fig:temprdecay}
\end{figure}

We now investigate the evolution of the properties of the thermal plasma at the East ribbon and coronal source compared to the full Sun spectral results. The presence of low-energy HXR from thermal bremsstrahlung from the ribbon sources is shown by imaging spectroscopy results, and can also be seen directly in the reconstructed images. HXR 4--9 keV emission associated with both ribbons (and coronal source) is shown in Figure \ref{fig:imsp_roi}b. We obtained images using two different imaging algorithms, Clean and the Maximum Entropy Method of the New Jersey Institute of Technology \citep[MEM-NJIT:][]{SchmahlPernakHurford:2007}, to confirm that the sources were not spurious artifacts of the algorithms. MEM-NJIT has a tendency to over resolve the sources, but in this case it works to confirm the overall spatial distribution of the low-energy HXR emission over part of the ribbons. We then constructed Clean maps (with \verb!Clean_beam_width!\footnote{Arbitrary factor applied to the beam convolved with Clean sources. Values between 1.5--2.0 tend to give better results when compared to different imaging algorithms \citep[see ][ and references therein.]{SimoesKontar:2013} } = 1.5; detectors 3--8) for nine time bins in the interval 06:22:22 and 06:27:51~UT. The spectra were fitted with an isothermal plus thick-target component. The latter was included to achieve a better fitting for the temperature, as fluxes around 8--10 keV and above could not be fitted with a thermal component only, requiring a non-thermal tail. In Figure \ref{fig:spot_em_t} we show the EM and $T$ for the East ribbon (orange), coronal source (blue) and full Sun (grey). Although the uncertainties in the imaging spectroscopy method are larger due to the smaller time intervals considered, the temperature peak around 06:25:40~UT is evident for both East ribbon and coronal source, in good agreement with the full Sun values from both RHESSI and GOES (see Figure \ref{fig:overview}d). {The evolution of the plasma temperature in the impulsive phase of this event is in fact quite unusual. The plasma temperature from GOES and RHESSI (Figure \ref{fig:temprdecay}) show that the temperature peak (06:25:38~UT) occurs about 8 to 10 seconds\footnote{Given the time resolution of GOES ($\approx$ 2 $s$) and RHESSI ($\approx$ 4 $s$)} before the HXR peak (06:25:46~UT). Even considering the different instrumental responses from GOES and RHESSI, both sample the hottest plasma present in flares, weighted by the EM \citep{RyanOFlannagainAschwanden:2014}, indicating that this temperature peak reveals an impulsive heating followed by a fast cooling process. The temperature decrease from 13 MK to 11MK in about two minutes is consistent with conduction losses, with a time scale of about 14 to 38 seconds \citep[for the coronal source,][]{Simoes:2015}. The $e$-folding cooling time for the (full Sun) temperatures in Figure \ref{fig:temprdecay} is $\tau=$ 35\,--\, 55 seconds. }

\begin{figure} 
\centerline{\includegraphics[angle=0,width=\textwidth]{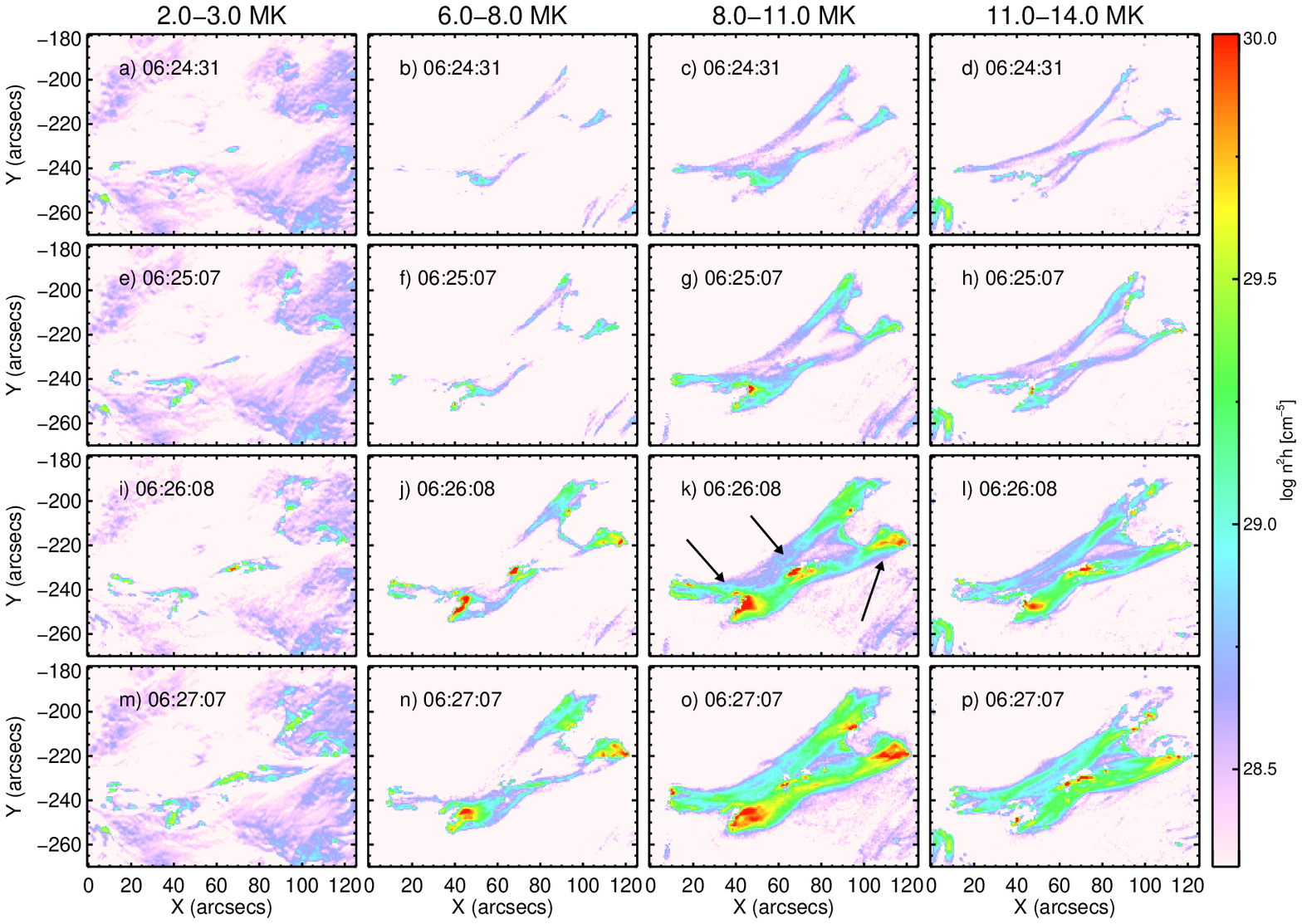}}
\caption{EM maps obtained from the AIA-DEM method \citep{HannahKontar:2012}, for temperature ranges 2--3, 6--8, 8--11 and 11--14 MK, at four times during the impulsive phase. The arrows (from left to right) in panel {\em k} indicate the positions of the East ribbon, coronal source and West ribbon.}
\label{fig:demmap}
\end{figure}

Since the spatial resolution of RHESSI is  lower than that of AIA, we confirm the location of the highest EM by applying the AIA-DEM method for each pixel in AIA images, following \cite{HannahKontar:2013}. In Figure \ref{fig:demmap} we show EM maps from the AIA-DEM method, where the largest concentrations ($\mathrm{EM}_c > 10^{29.5}$ cm$^{-5}$) of hot plasma ($T>8$ MK) are near the flare ribbons and the coronal source.  

EM maps (Figure \ref{fig:demmap}) obtained with the AIA-DEM method also suggest the presence of short-lived, high-EM ribbon ($\mathrm{EM}_c \approx 10^{30}$ cm$^{-5}$) at 11-14 MK near the HXR peak time at 06:25:46~UT. 

The gradual rise of the EM values from the AIA-DEM (Figure \ref{fig:em_time}) starting about 06:24~UT shows the filling of the loops with hot material but the sudden enhancement between 06:25:46 and 06:25:55~UT comes from this impulsively heated material, with the two components then contributing to the overall EM. The fast cooling of the 13 MK source can only be attributed to conduction to the lower atmosphere, as radiative cooling is much slower than the observed $e$-folding time, of about 35\,--\,55 seconds. It is not clear what is the source of energy to produce the temperature peak, however we here speculate a scenario where a rise in the plasma temperature leads the acceleration of particles.

\subsection{Energy Budget at the 10 MK Ribbon}\label{sec:heating}

We now examine the energy budget of the 10 MK plasma in the East ribbon. For that, we consider a slab of plasma at $T=10$ MK in the lower atmosphere (\textit{i.e.} at transition region heights), with a thickness $L$, being collisionally heated by the non-thermal electrons, balanced by radiative and conductive losses. 

Following \cite{FletcherHannahHudson:2013} we calculate the energy budget of the 10 MK plasma, considering radiative and conductive losses and the collisional energy loss of the non-thermal electrons as the energy input mechanism. The total collisional power input $P_{\mathrm{coll}}$ [erg s$^{-1}$] to the plasma can be inferred from the collisional thick-target model by integrating the energy loss \citep{Emslie:1978,FletcherHannahHudson:2013}

\begin{equation}
P_{\mathrm{coll}}=F_\mathrm{tot}E_c\frac{\delta-1}{\delta-2}\left[1-\left(\frac{\delta}{2}-1 \right)x_c^{1-\delta/2}B(x_c;\frac{\delta}{2}-1;\frac{3}{2}) \right],
\end{equation}
where $B$ is the incomplete beta function and $x_c={3KN}/{E_c^2}$, where $K=2\pi e^4$, and also
\begin{equation}
F_\mathrm{tot}=\int_{E_c}^\infty F_0E_0^{-\delta}dE_0
\end{equation}
being $F_\mathrm{tot}$ the total number of electrons per second above $E_c$ injected into the source.

The values for $F_{tot}$, $\delta$ and $E_c$ are taken from RHESSI spectral analysis (Figure \ref{fig:rhessi}). In order to estimate the time evolution of these parameters for the East ribbon source, we used the results from the full Sun spectroscopy and assumed that about a third of the total number of electrons are associated with each of the three HXR sources, based on the HXR imaging spectroscopy results in Table \ref{tab:imsp}.

The column depth, $N$, can be estimated from the column, EM$_c$, inferred with the AIA-DEM analysis, as $N = n_eL \simeq (\mathrm{EM}\ L)^{1/2}$, for a uniform source of thickness $L$.

The hot plasma will lose energy by conduction, along the field direction $z$, to cooler layers of the atmosphere at a rate $L_\mathrm{cond}$ [erg s$^{-1}$]. We estimate the conductive losses following \cite{BattagliaFletcherBenz:2009}, considering the case of flux-limited conduction: 
\begin{equation}
L_{\mathrm{cond}}=\varrho(x)\kappa_0T^{5/2}\frac{dT}{dz}A
\label{eq:cond}
\end{equation}
where the factor $\varrho(x)$, a function of $x=\log(l_\mathrm{mfp}/L)$, reduces the classical \cite{Spitzer:1965} conduction coefficient $\kappa_0=10^{-6}$ erg cm$^{-1}$ s$^{-1}$ K$^{-7/2}$ to the flux-limited conduction \citep{Campbell:1984}. \cite{BattagliaFletcherBenz:2009} fitted the values of $\varrho$ published by \cite{Campbell:1984} as $\varrho (x)=1.01 \mathrm{e}^{-0.05(x+6.63)^2}$. This condition applies because the chromospheric scale length is small and the electron collisional mean free path, $l_\mathrm{mfp}=5.21\times 10^3 T^2/n_e$, is significant compared to the temperature scale length. Here, we approximate $dT/dz$ by $T/L$ and set $T=10$ MK and $L=2000$ km.

The optically thin hot plasma radiates energy at a rate $L_\mathrm{rad}$ that can be estimated by \citep{RosnerTuckerVaiana:1978}:
\begin{equation}
L_\mathrm{rad}=10^{-17.73}\mathrm{EM}~T^{-2/3}, (10^{6.3}<T<10^7~\mathrm{K}).
\end{equation}

\begin{figure} 
\resizebox{\hsize}{!}{\includegraphics[angle=0]{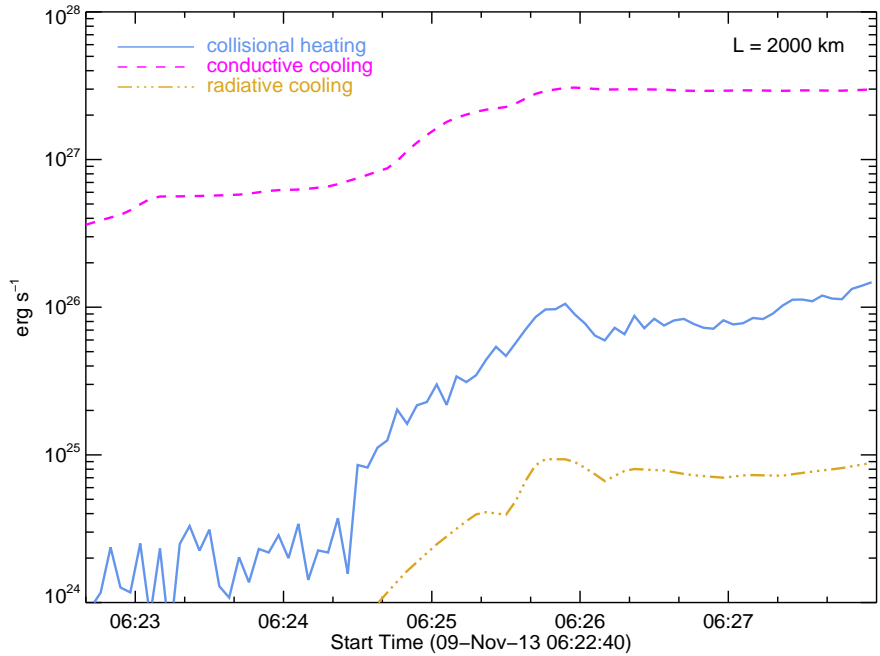}}
\caption{Power budget of the East ribbon source, considering beam collisional heating (blue), conductive (magenta) and radiative (gold) cooling, in a $L_T=2000$ km thick slab of plasma at a temperature of 10 MK.}
\label{fig:power}
\end{figure}

Using the EM values derived from the AIA-DEM analysis (Figure \ref{fig:em_time}c), and the area $A$ of the 10 MK source estimated as defined in Section \ref{sec:dem}, we found that the collisional heating is not sufficient to balance the estimated conductive losses as shown in Figure \ref{fig:power}. 

\cite{FletcherHannahHudson:2013} found a substantial amount of plasma at 10 MK in the flare ribbons during the early impulsive phase of the flare SOL2010-08-07T18:24, with an average column EM of a few times 10$^{28}$ cm$^{-5}$. They found that the energy carried by an electron beam is not sufficient to heat the ribbon plasma when radiative and conductive losses are taken into account, unless a low-energy cutoff of $E_c \approx 5$ keV is considered. The total electron energy inferred from the HXR spectrum is mostly defined by the low-energy cutoff $E_c$ \citep[\textit{e.g.}][]{Saint-HilaireBenz:2005}, which is sometimes chosen arbitrarily to balance the energy deposited by the non-thermal electrons and the value of the maximum thermal energy \citep[\textit{e.g.}][]{MrozekTomczak:2004}. The $E_c$ values we found from the HXR spectral fitting are already quite low, approximately $8$ keV during the impulsive phase (see the bottom panel in Figure \ref{fig:rhessi}), only a few times the average thermal energy $kT$ of electrons in the 10 MK plasma. { It is of course possible to reduce the value of the low-energy cutoff to make the energy losses and gains balance, and a lower value of the low-energy cutoff is permitted by the HXR spectrum. A lower $E_c$ means an increased total `non-thermal' power, and an increased electron number flux.  For this event, a value of $E_{c}\approx 4$ keV is necessary to equate energy gains and losses.} On the other hand, using the $E_c$ values obtained from the HXR spectral analysis, a slab thickness $L$ of $\approx$30 Mm is required to balance the energy budget, which would comprise a large portion of the loops, a picture not supported by the AIA images, which show very compact footpoint sources.  

As shown by \cite{Simoes:2015}, the coronal source is both dense ($n \approx 10^{11}$cm$^{-3}$) and hot ($T\approx 13$ MK), a downward thermal conduction from the coronal source to the ribbons may contribute to the observed heating at these regions \citep[\textit{e.g.}][]{HoriYokoyamaKosugi:1998,QiuSturrockLongcope:2013,BattagliaFletcherBenz:2009}. 

{ If the low energy cutoff is indeed as low as 4 keV to balance energy gains and losses by the chromospheric plasma then  $E_c$ approaches the mean thermal energy of the 10 MK plasma, and it becomes difficult to separate the `beam' from the thermal core. We may find it profitable in the future to think about the electrons in the chromosphere forming a single distribution, such as a $\kappa$-distribution, which describes a single population transitioning smoothly from a power-law-like tail to a thermal-like core. This distribution is being discussed in connection with dense coronal HXR sources, where populations of heated electrons and accelerated electrons are present in the same volume \citep{Oka:2013,Oka:2015}. \cite{Bian:2014} have demonstrated how a $\kappa$-distribution arises naturally in a volume where diffusive acceleration and collisional losses operate simultaneously and co-spatially. If this volume is coronal and a beam is formed by a high energy tail that `leaks out'  then that beam looks very much like a standard power-law beam, with a low energy cutoff determined by the particle velocity at which the escape time is less than the acceleration or diffusion timescales; then, we are then still in the standard collisional thick-target scenario but with a slightly differently shaped coronal beam. However, if we move the site of both heating and acceleration to the chromosphere, which is a very different interpretation, then the properties of the $\kappa$-distribution extending across all of the electron energy space can be investigated to evaluate the total required power and number flux needed to account for all radiation signatures. This is an interesting possibility; it does however require that some other agent, such as wave turbulence, is present in the chromosphere to locally heat and accelerate electrons. A full investigation is beyond the scope of this paper.}

\section{Summary}

We have presented an analysis of the plasma in the flare ribbons of the event SOL2013-11-09, a C2.6 class non-eruptive, two-ribbon flare using SDO/AIA EUV and RHESSI HXR observations. The ribbons have impulsive EUV/UV emission seen in all SDO/AIA filters, well associated with non-thermal HXR emission observed by RHESSI. Using the method of regularised inversion of SDO/AIA data \citep{HannahKontar:2012} we obtained the differential emission measure (DEM) of the two flare ribbons, and investigated the time evolution of the emission measure (EM) in three temperature ranges (0.5 -- 1.4, 1.8 -- 3.2 and 7.9 -- 12.6 MK). From these, we have shown that the plasma heats rapidly to 12--13 MK during the impulsive phase of the event, marked by the HXR peak. The EM temporal evolution shows a peak near the time of maximum HXR emission, indicating fast heating and cooling with the hottest plasma ($T=$7.9\,--\, 12.6 MK) reaching EM values of 1 \,--\, 3 $\times 10^{47}$ cm$^{-3}$, these values agreeing with those obtained from GOES and RHESSI. The rapid evolution of the ribbon plasma temperature and its high peak temperature are confirmed by RHESSI imaging spectroscopic analysis, and also agree with the temperature derived from GOES (although without spatial resolution). Also we note that the evolution of the ribbon plasma characteristics are very similar to the those found in the intense and compact coronal source, as studied in detail by \cite{Simoes:2015}. Performing RHESSI HXR imaging spectroscopy, we obtained the parameters to describe the distribution of non-thermal electrons at each source (both ribbons and coronal source). With the information about the plasma and non-thermal electrons at the East ribbon, we deduced the energy balance of the plasma, considering collisional beam heating \citep{Emslie:1978,FletcherHannahHudson:2013} against conductive and radiative losses. We found that beam heating alone is not sufficient to heat and maintain the ribbon plasma at $T=10$ MK, even with a low energy cutoff of $E_c\approx 8$ keV, and speculate if the dense and hot coronal source can provide a heat source for the ribbons by conduction \citep[\textit{e.g.}][]{HoriYokoyamaKosugi:1998,QiuSturrockLongcope:2013}.

\begin{acks}
The authors would like to thank Paul Boerner for providing updated SDO/AIA temperature response functions and Iain Hannah for making the DEM regularised inversion software freely available. The research leading to these results has received funding from the European Community’s Seventh Framework Programme (FP7/2007- 2013) under grant agreement no. 606862 (F-CHROMA), from STFC grant ST/I001808/1 (PJAS, LF) and ST/L000741/1 (LF), and from an STFC ‘STEP’ award to the University of Glasgow (DRG). 
\end{acks}

\bibliographystyle{spr-mp-sola}
\bibliography{refs}
\end{article} 
\end{document}